\documentclass[twocolumn,superscriptaddress,longbibliography,aps,pra,preprintnumbers,10pt]{revtex4-2}
\usepackage[utf8]{inputenc}
\usepackage[urlcolor=blue,colorlinks=true,citecolor=blue,linkcolor=blue,pdfstartview={FitH},bookmarks=false]{hyperref}
\usepackage[T1]{fontenc}
\usepackage{graphicx}
\usepackage{xcolor}
\usepackage{amsmath,amssymb,amsfonts}

\newcommand{\sizeone}{1.0\textwidth}
\newcommand{\sizetwo}{0.497\textwidth}

\newcommand{\ext}{pdf}

\begin{document}

\title{Localized fermions on the triangular lattice with Ising-like interactions}

\author{Lubom\'ira Regeciov\'a}
\email[e-mail: ]{regeciova@saske.sk}
\homepage[\mbox{ORCID ID}: ]{https://orcid.org/0000-0002-9154-1878}
\affiliation{Institute of Experimental Physics Slovak Academy of Sciences, Watsonova 47, 040 01 Ko\v{s}ice, Slovakia}

\author{Konrad Jerzy Kapcia}
\email[e-mail: ]{konrad.kapcia@amu.edu.pl}
\homepage[\mbox{ORCID ID}: ]{https://orcid.org/0000-0001-8842-1886}
\affiliation{\mbox{Institute of Spintronics and Quantum Information, Faculty of Physics and Astronomy}, Adam Mickiewicz University in Pozna\'n, 
ul. Uniwersytetu Pozna\'{n}skiego 2, 61614 Pozna\'{n}, Poland}

\date{\today}

\begin{abstract}
The model of localized fermions on the triangular lattice is analyzed in means of the Monte Carlo simulations in the grand canonical ensemble.
The Hamiltonian of the system has a form of the extended Hubbard model (at the atomic limit) with nearest-neighbor Ising-like magnetic $J$ interactions and onsite Coulomb $U$ interactions.
The model is investigated for both signs of $J$, arbitrary $U$ interaction and arbitrary chemical potential $\mu$ (or, equivalently, arbitrary particle concentration $n$).
Based on the specific heat capacity and sublattice magnetization analyses, the phase diagrams of the model are determined.
For ferromagnetic case ($J<0$), the transition from the ordered phase (which is a standard ferromagnet and can be stable up to $k_{B}T/|J| \approx 0.61$) is found to be second-order (for sufficiently large temperatures $k_{B}T/|J| \gtrsim 0.2$) or first-order (for $-1<U/|J|<-0.65$ at the half-filling, i.e., $n=1$).
In the case of $J>0$, the ordered phase occurs in a range of $-1/2<U/|J|<0$ (for $n=1$), while for larger $U$ the state with short-range order is also found (also for $n \neq 1$).
The ordered phase is characterized by an antiferromagnetic arrangement of magnetic moments in two sublattices forming the hexagonal lattice.
The transition from this ordered phase, which is found also for $\mu \neq 0$ ($n \neq 1$) and $U/|J|>-1/2$  is always second-order for any model parameters.
The ordered phase for $J>0$ can be stable up to $k_{B}T/|J| \approx 0.06$.
\end{abstract}

\maketitle

\section{Introduction}

Systems with geometrical frustration represent a fascinating area of condensed matter physics, where the arrangement, for example, of magnetic moments leads to an inability to simultaneously satisfy all interactions within the system \cite{Diep2013frustrated,StarykhRPP2015}. 
This phenomenon is exhibited in systems where the geometry of the lattice creates competing interactions, such as those found in triangular or tetrahedral arrangements of the magnetic moments~\cite{ZukovivcAPPA2010,BishopIJMPB2011,LiPLA2011,StreckaAPS2015}

The Ising model $\hat{H}_{I} = \sum_{i,j}J_{i,j} \hat{s}_i \hat{s}_j$ is one of the simplest model which can describe the magnetism in the system of localized magnetic moments.
Its exact solutions with interactions restricted to the nearest-neighbors (i.e., $J_{ij} \equiv J/z$ if $i$ and $j$ are neighboring site and $J_{ij} \equiv 0$ otherwise) for the one-dimensional chain as well as the two-dimensional square lattice for both signs of the interaction are very well-known \cite{Ising1925,OnsagerPR1944,KaufmanPR1949}.
The model on the triangular lattice has been also extensively studied. 
As that lattice has a geometrical frustration, the system exhibits different behaviors, which are dependent on the sign of interaction $J$.
Rigorous results for this lattice \cite{HoutappelPhys1950a,HoutappelPhys1950b} show that the transition temperature is $k_BT_c/|J| = 0.607$ for ferromagnetic case ($J<0$), whereas for antiferromagnetic case ($J>0$) the transition is absent and there is no long-range order in the system at any temperature. 
Within the use of Bethe-Peierls approximation \cite{CampbellPRA1972} and Monte Carlo simulations \cite{MihuraPRL1977}, it was shown that ordered phase (with so-called $\sqrt{3} \times \sqrt{3}$ order) can occur in the model with $J>0$ for non-zero fields or for large enough next-nearest-neighbor interactions in finite temperatures (cf. also mean-field solutions \cite{KincaidPR1975}).
With longer-range interactions even more complex ordered patterns are possible \cite{MetcalfPLA1974,KaburagiJJAP1974,KaburagiJPSJ1978}.

On the other hand, for the description of strongly correlated electron systems, or more generally - fermionic systems, the Hubbard model $ \hat{H}_H = t \sum_{\langle i,j \rangle,\sigma} \hat{c}_{i\sigma}^{\dag} \hat{c}_{j\sigma}^{\ } + U \sum_{i} \hat{n}_{i\uparrow} \hat{n}_{i\downarrow} $ was introduced \cite{HubbardPRSL1963,PennPR1966}.
This model captures the itinerant nature of the fermions via hopping term (with $t$) and Mott localization (and/or superconductivity) due to Coulomb repulsion (effective attraction, respectively) $U$ between the particles \cite{Spalek1987,GeorgesRMP1996,ImadaRMP1998,KotliarRMP2006,GeorgescuRMP2014,MicnasRMP1990}.
Inclusion of different types of intersite interactions between fermions leads to effective descriptions of various exotic phenomena, for instance, various charge-orderings or intersite superconductivity \cite{MicnasRMP1990,SpalekAPPA2007,DuttaRPP2015}.

In this work, motivated by a very complex structure of phase diagrams of the model of charged particles with the repulsive intersite interaction on the triangular lattice \cite{KapciaJSNM2019,KapciaNano2021,KapciaJMMM2022},  we investigate the model of the localized fermions interacting via spin-spin interactions between nearest-neighbor sites as well as on-site density-density interaction.
More specifically, we study the extended Hubbard model with intersite Ising-like magnetic interactions at the atomic limit on the triangular lattice with the following form: 
\begin{equation}\label{eq:hamiltonian}
    \hat{H} = U \sum_i \hat{n}_{i\uparrow} \hat{n}_{i\downarrow} + \frac{2J}{z} \sum_{\langle i,j\rangle}\hat{s}_i \hat{s}_j  - \left(\mu + \frac{U}{2}\right) \sum_i \hat{n}_i,
\end{equation}
where $\hat{c}_{i\sigma}^{\dag}$  ($\hat{c}_{i\sigma}^{\ } $) denotes the creation (annihilation) operator of a fermion with spin $\sigma$ ($\sigma \in \{ \uparrow, \downarrow \}$) at lattice site $i$ and other operators are defined as
$\hat{s}_i = \left( \hat{n}_{i\uparrow} - \hat{n}_{i\downarrow} \right)/2$,
$\hat{n}_i =  \hat{n}_{i\uparrow} + \hat{n}_{i\downarrow}$,
$\hat{n}_{i\sigma} = \hat{c}_{i\sigma}^{\dag} \hat{c}_{i\sigma}^{\ } $.
$\sum_{\langle i, j \rangle}$ is a summation over nearest-neighbor sites independently and $z$ is a coordination number (number of the nearest neighbors). 
$U$ denotes the onsite Hubbard interaction and $J$ is the Ising-like magnetic interaction between nearest-neighbor sites $i$ and $j$. 
Finally, $\mu$ stands for the chemical potential (shifted by $-U/2$), which controls total particle concentration $n$ in the system (for $\mu=0$ one gets the half-filling $n=1$)~\cite{KapciaJMMM2024,KapciaPhysA2015}.

Note that model (\ref{eq:hamiltonian}) is equivalent with the aforementioned Ising model (in the absence of the external magnetic field) then and only then if $U\rightarrow+\infty$ and $\mu=0$ in Eq. (\ref{eq:hamiltonian}) (the half-filling case) \cite{KapciaPhysA2015,KapciaJMMM2024}.
In such a limit of model (\ref{eq:hamiltonian}) only two states: (i) a particle with spin-$\uparrow$ $|\uparrow \rangle_i$ or (ii) a particle with spin-$\downarrow$ $|\downarrow \rangle_i$ are possible at every lattice site.
Thus, model (\ref{eq:hamiltonian}) can be considered as a generalization of the Ising model for fermionic particles (with a variable number of the particles). 

\begin{figure*}[bth]
\includegraphics[width=\sizetwo]{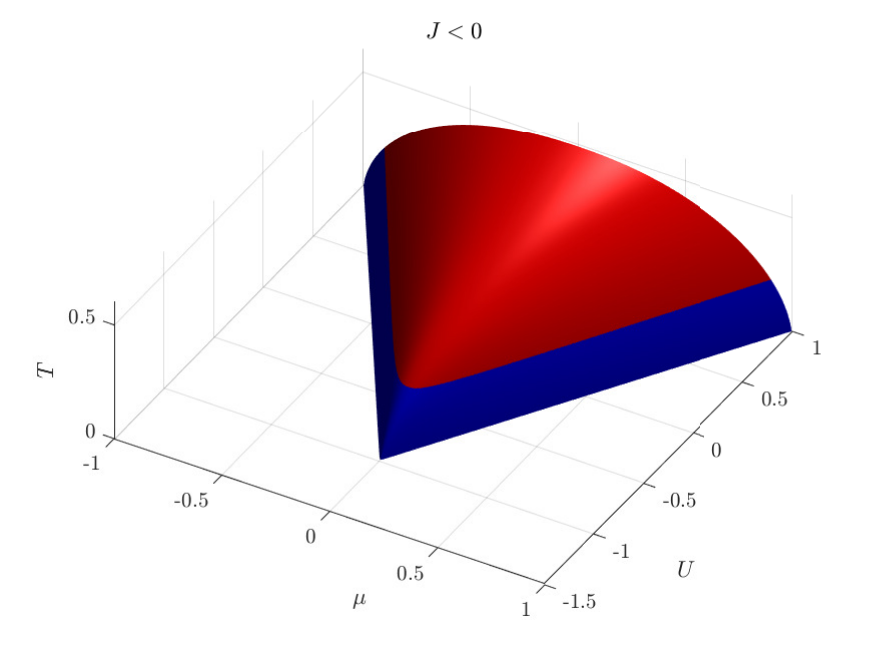}
\includegraphics[width=\sizetwo]{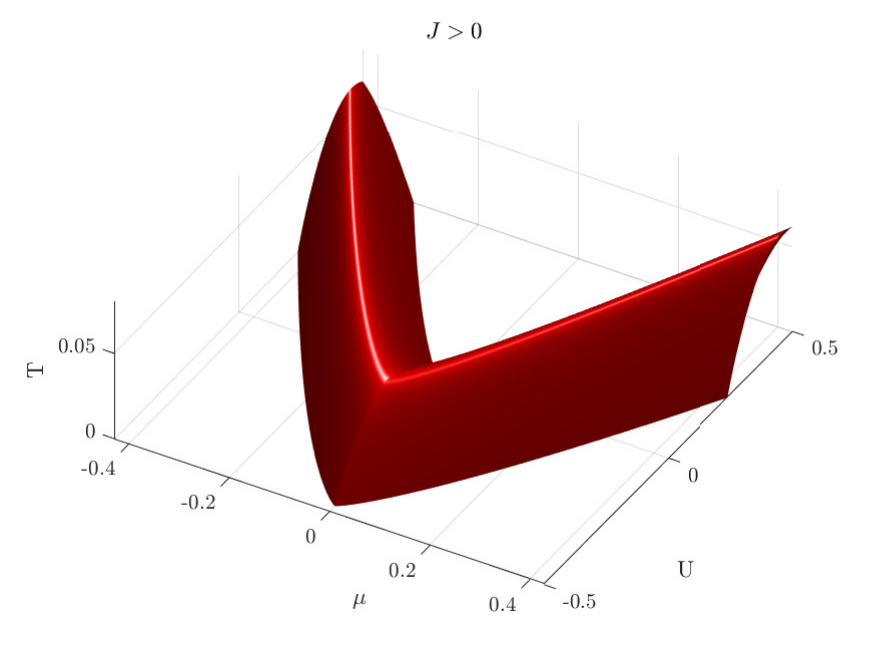}
\caption{\label{fig:propfd}%
Schematic three dimensional phase diagrams (in $\mu$-$U$-$k_{B}T$ space) for ferromagnetic case ($J<0$, the left panel) and antiferromagnetic case ($J>0$, the right panel). The blue and red surfaces denote first order (discontinuous) and second order (continuous) transitions.}
\end{figure*}

Model (\ref{eq:hamiltonian}) was extensively studied for hyper-cubic lattices. 
In such a case there is a full correspondence between $J<0$ and $J>0$ cases. 
In the variational approach with the mean-field approximation for the intersite term and rigorous treatment of the single-site term   \cite{RobaszkiewiczPSSB1975,RobaszkiewiczAPPA1979,KlobusAPPA2010,KapciaPhysA2015} (rigorous solution for $z\rightarrow +\infty$), the phase transition between magnetically ordered phases (ferromagnetic or antiferromagnetic depending on the sign of $J$ interaction) was found to be first-order (discontinuous) for $k_BT/|J|<1/3$ and second-order (continuous) for $k_BT/|J|>1/3$. 
The tricritical point, where transition changes its order at half-filling is located at $U/|J| = (2/3)\ln 2$ \cite{RobaszkiewiczAPPA1979,KapciaPhysA2015}.
The Monte Carlo simulations were performed for the model on two-dimensional square lattice with $z=4$ \cite{MurawskiAPPA2012,MurawskiAPPA2014,MurawskiAPPA2015}. 
It was found that the structure of the phase diagrams is similar as in the variational approach with reduction of the range of ordered phase occurrence, e.g., the transition temperatures are reduced if compared with $z\rightarrow + \infty$ limit.
Also some exact results for one-dimensional chain ($z=2$) were obtained \cite{ManciniOP2012,ManciniEPJB2013}.
Unfortunately, any rigorous and/or analytical results for model (\ref{eq:hamiltonian}) are not known in dimensions $1<d<+ \infty$ up to now (apart from those in the mean-field approximation).

The variational approach was also used to inspect the phase diagram of model (\ref{eq:hamiltonian}) on the triangular lattice ($z=6$) \cite{KapciaJMMM2024}. 
It was shown that  phase diagrams for both ferromagnetic and antiferromagnetic cases are similar. 
The only difference found is that the values of $k_BT/|J|$, $\mu/|J|$, and $U/|J|$ in ferromagnetic case should be reduced by factor $2$ to get the results in antiferromagnetic case. 
In this approach, the results in $J<0$ case are the same for both hypercubic and triangular lattices \cite{KapciaPhysA2015,KapciaJMMM2024}.

In this work, we investigate model (\ref{eq:hamiltonian}) on the triangular lattice in the means of the Monte Carlo method (for details see Sec. \ref{sec:method}). 
We present studies for both signs of $J$ interaction: $J<0$ (Sec. \ref{sec:ferro}) and $J>0$ (Sec. \ref{sec:antiferro}). 
From analysis of the temperature dependence of specific heat capacity and sublattice magnetizations, we find the ordered phases (including states with short-range order) and determine the phase diagrams for different chemical potentials (effectively for different concentrations as the concentration of particles $n$ in the system is determined by potential $\mu$).

\section{Method description: Monte Carlo algorithm}
\label{sec:method}

To explore the temperature-dependent properties of the model, we have used the classical Monte Carlo method with the Metropolis algorithm, a standard approach for such lattice-based systems \cite{Newman1999monte,binder2010monte,landau2021guide}. 
Our simulations are performed in the grand canonical ensemble, where the number of particles is not fixed, which permits fluctuations of the number of particles. 
We start at finite and sufficiently high temperature, $k_{B}T/|J| = 0.5$ (or $k_{B}T/|J| = 1.0$ for interactions $U/|J| > 1$), to ensure thorough exploration of the state space. 
The temperature is then gradually decreased down to a final temperature of $k_{B}T/|J| = 0.005$, which is sufficient to describe the ground-state properties.
In our implementation, for each temperature, the initial $10^6$ Monte Carlo steps are discarded for equilibrium consideration, and an additional $10^6$ steps are retained for statistical averaging in the simulation.

\begin{figure*}[tbh]
\includegraphics[width=\sizeone]{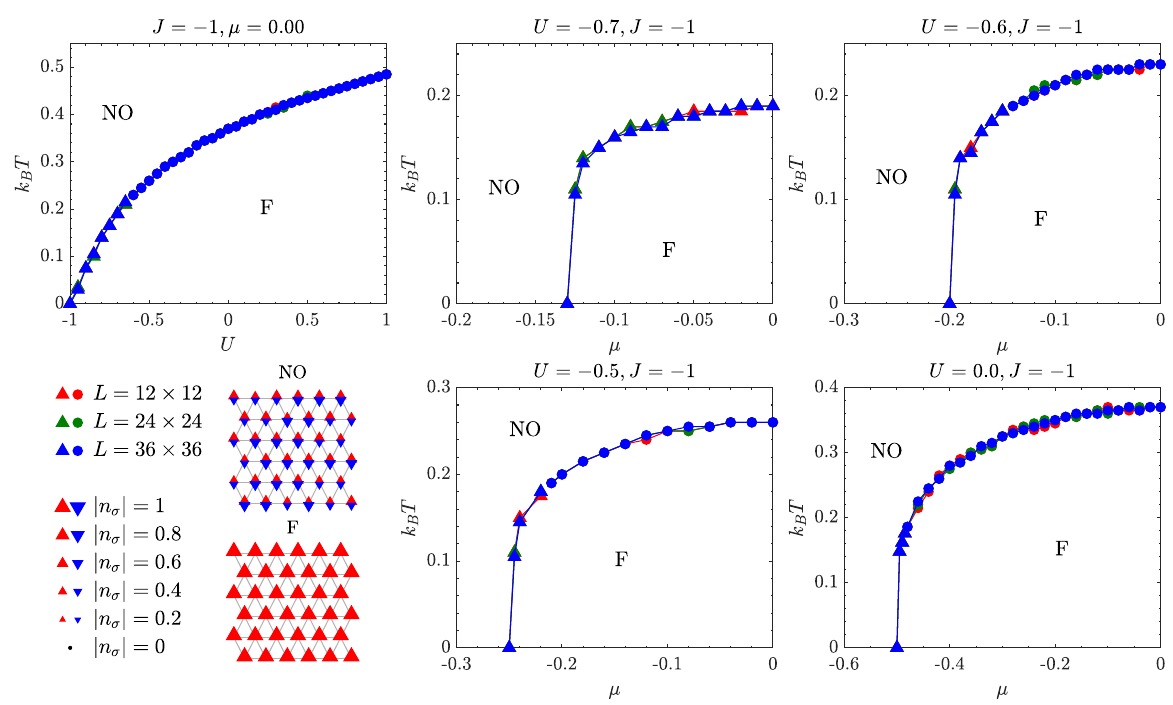}
\caption{\label{fig:PDferro}%
Phase diagrams for the model with ferromagnetic ($J=-1$) interactions, shown for three different lattice sizes ($L = 12^{2},24^{2}, 36^{2}$).
Triangle markers indicate first-order phase transitions, while circular markers represent second-order phase transitions.
In the bottom-left corner, Monte Carlo averaged spin configurations for the identified ferromagnetic (F) phase at $U=-0.6$, $\mu=0.0$, and $k_BT=0.15$ and non-ordered (NO) phase at $U=-0.6$, $\mu=0.0$, and $k_BT=0.4$, are shown for a lattice of size $L = 6 \times 6$.}
\end{figure*}

\begin{figure*}[t]
\includegraphics[width=\sizeone]{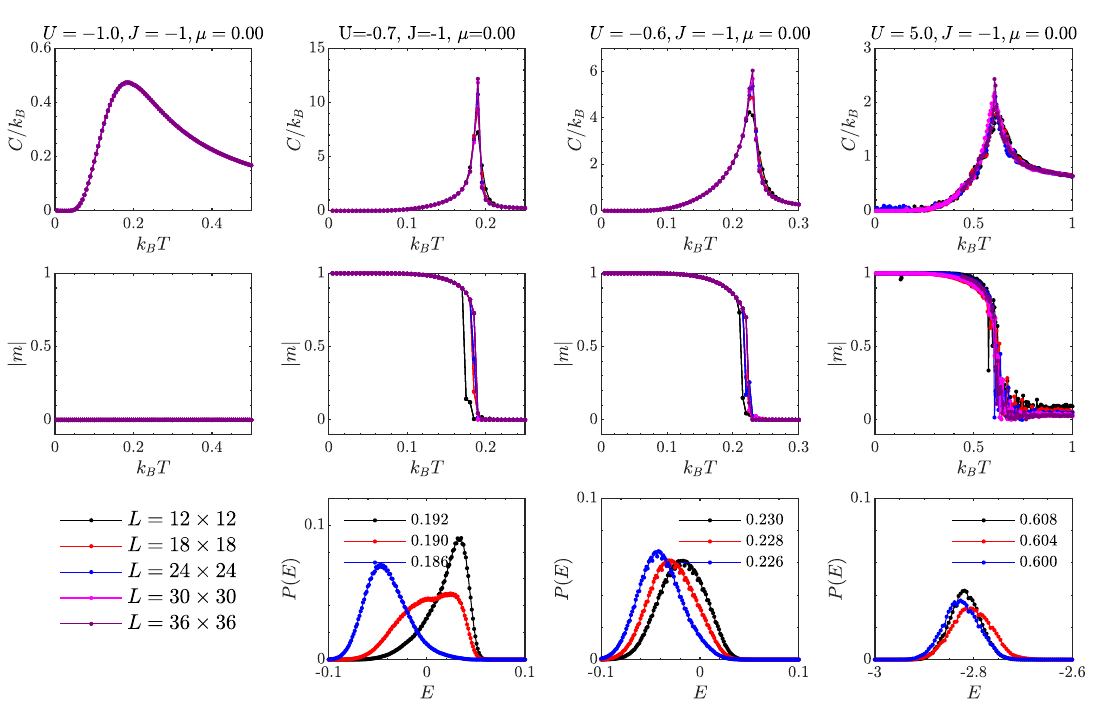}
\caption{\label{fig:propferathf}%
Specific heat capacities (top row), sublattice magnetizations (second row), and the energy distribution functions near the phase transition on a $24 \times 24$ lattice (third line) for ferromagnetic interaction ($J=-1$) and representative values of $U$ ($U = -1.0, -0.7, -0.6, 5.0$, from the left) at $\mu=0$.}
\end{figure*}

Our triangular clusters consist of $L$ lattice sites, each of which can occupy by one of four possible states, representing the presence or absence of the particle with spin-$\uparrow$ or spin-$\downarrow$.
More specifically, every lattice site $i$ can be occupied by (i) a particle with spin-$\uparrow$: $|\uparrow \rangle_i$, (ii) a particle with spin-$\downarrow$: $|\downarrow \rangle_i$, (iii) two particles, one with spin-$\uparrow$ and one with spin-$\downarrow$ (double occupied): $|\uparrow \downarrow \rangle_i$, or (iv) the site can be empty (no particles at the site): $|0\rangle_i$.
Note that this lattice can be divided into three equivalent sublattices  \cite{CampbellPRA1972,StarykhRPP2015,KapciaNano2021,KapciaJMMM2022,KapciaJMMM2024}.
The study of this strongly frustrated system is computationally intensive, especially for the case of antiferromagnetic $J$, which limits the size of clusters we can effectively study. 
For this reason, we study triangular clusters of size up to $L = 36 \times 36$, which are sufficient to provide reliable analysis of finite-size effects and to extrapolate the results to the thermodynamic limit.
A similar Monte Carlo approach is widely used to explore temperature-dependent properties and phase behavior in strongly correlated electronic systems \cite{Pawlowski2006,Ganzenmuller2008}.

\section{Results and discussion}
\label{sec:results}

In this section, the results of the simulations for both signs of the intersite magnetic interaction are presented. 
We present the phase diagrams, which are determined by careful examination of the temperature dependence of specific heat capacity $C = d E/ dT$ ($E$ - the energy of the system per lattice site) and sublattice magnetization $m_{\alpha}$. 
In particular, the specific heat capacity is determined using the formula $C=\langle \varepsilon^2 \rangle - \langle \varepsilon \rangle ^2 / T^2$, where $\langle  \varepsilon \rangle$ 
($\langle  \varepsilon ^2 \rangle$) represents the averages of energies $\varepsilon$ ($ \varepsilon ^2$) sampled by the Monte Carlo algorithm.
Moreover, magnetization $m_\alpha$ is an average magnetic moment (per site, multiplied by the factor of $2$) in $\alpha$ sublattice ($\alpha \in \{ A,B, C\}$), which build the triangular lattice, cf., e.g., \cite{CampbellPRA1972,StarykhRPP2015,KapciaNano2021,KapciaJMMM2022,KapciaJMMM2024} ($m_{\alpha} = (6/L) \sum_{i \in \alpha} \langle \hat{s}_ i \rangle$).

\begin{figure*}[tbh!]
\includegraphics[width=\sizetwo]{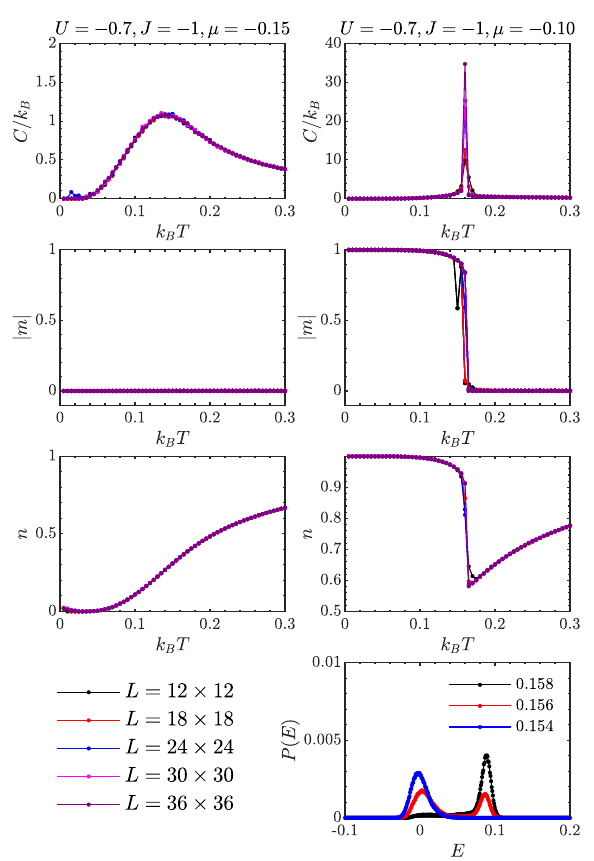}
\includegraphics[width=\sizetwo]{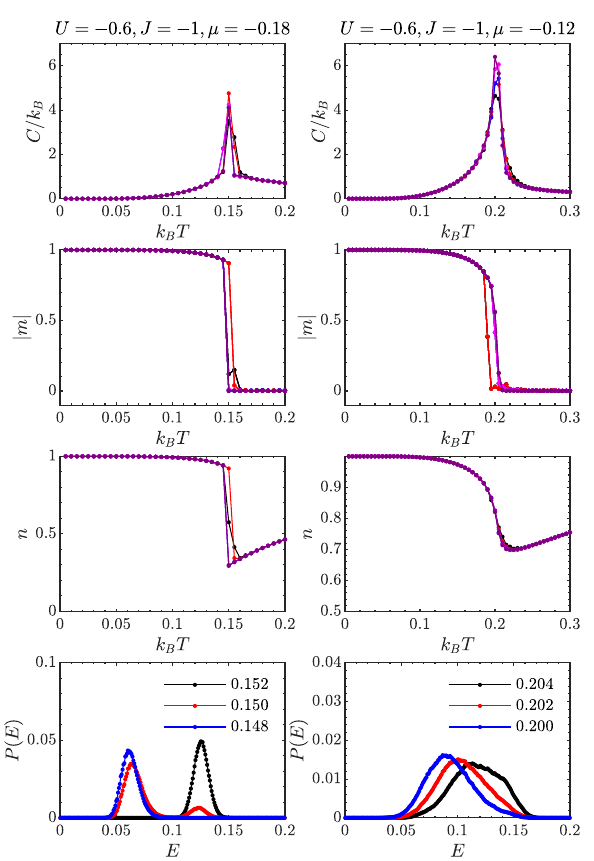}
\caption{\label{fig:propferawayhf}%
Specific heat capacities (top row), sublattice magnetizations (second row), number of particles (third row), and the energy distribution functions near the phase transition on a $24 \times 24$ lattice (fourth line) for ferromagnetic interaction ($J=-1$) and representative values of $U$ and $\mu$ (from the left: $U = -0.7$ and $\mu = -0.15, -0.10$; $U = -0.6$ and 
$\mu = -0.18, -0.12$).}
\end{figure*}

In the present work, all energies (i.e., $k_{B}T$, $U$, and $\mu$) are shown in units of $|J| = z \times (|J|/z)$ (the notation of Eq. (\ref{eq:hamiltonian})) in the full correspondence to the results previously presented in Refs. \cite{RobaszkiewiczPSSB1975,RobaszkiewiczAPPA1979,KlobusAPPA2010,KapciaPhysA2015,MurawskiAPPA2012,MurawskiAPPA2014,MurawskiAPPA2015,KapciaJMMM2024}.
Thus, in principle, we use $|J|=1$.
Because the model exhibits the particle-hole symmetry \cite{KapciaPhysA2015,KapciaJMMM2024}, $k_{B}T$-$\mu$ phase diagrams are symmetric with respect to $\mu=0$ and they are presented  only for $\mu \leq 0$ (or, equivalently, for $n \leq 1$).
Fig.~\ref{fig:propfd} shows schematic phase diagrams of model (\ref{eq:hamiltonian}) for both signs for $J$ interaction and illustrates the positions of phase boundaries of ordered phases (defined in the next section) of first (red) and second (blue) order in the $\mu-U-T$ space.
In the figure for $J>0$ case, the region of the occurrence of the short-range order inside the non-ordered phase is not indicated.

\subsection{Ferromagnetic interactions}
\label{sec:ferro}

First, we discuss the case of ferromagnetic intersite interactions (i.e., $J<0$). 
In this case, there is no frustration in the system, and we have identified two phases occurring in it: (i) the non-ordered (NO) phase which is defined by zero average magnetization in each sublattice ($m=m_{\alpha}=0$),
indicating equal average occupancies of spin-$\uparrow$ and spin-$\downarrow$ particles and the absence of both long-range and significant short-range magnetic order
and (ii) the ferromagnetic (F) phase in which $m=m_{\alpha}$ and $|m|>0$ (all three sublattices have the same magnetization).
In both phases, concentrations of the particles in each sublattice are the same, i.e., $n=n_{\alpha}$.
Note that, in this case, the spin configurations with $m=|m|$ (all averaged moments $\uparrow$-direction) and $m=-|m|$ (all all averaged moments $\downarrow$-direction) are equivalent (because there is no external magnetic field in the system).

Let us begin our discussion with the study of the influence of the Hubbard interaction $U$ at zero chemical potential $\mu=0$ (which corresponds to the half-filling in model (\ref{eq:hamiltonian}), $n=1$), for which the phase diagrams are depicted in Fig.~\ref{fig:PDferro} (the left top panel). 
One can see that the ferromagnetic interaction $J$ leads to a relatively simple $k_{B}T$--$U$ phase diagram with a phase transition from the non-ordered phase (NO) to the ordered (F) phase for $U>-1$, moving towards higher temperatures $k_BT$ with increasing $U$. 
At sufficiently large on-site density interaction $U$ ($U > 3$), it reaches an almost constant value of $k_BT \approx 0.605$, which aligns with the exact solution of the classical Ising model, to which our model can be mapped in the limit of large $U$ (for $\mu=0$).

These phase transitions are obtained from the analysis of specific heat capacity and sublattice magnetizations on finite lattices with $L = 12 \times 12$ to $L = 36 \times 36$ sites, which are illustrated in Fig.~\ref{fig:propferathf} for representative values of $U$ ($U=-1.0,-0.7,-0.6,5.0$). 
One can see that in the case of $U \leq -1$ (the NO phase), we have found a broad maximum in the specific heat capacity curves which does not change with increasing $L$, and there is also no change in sublattice magnetizations, which equal zero. 
This can be attributed to the presence of a Schottky anomaly, which indicates a small number of discrete energy levels that dominate the system behavior and quantify the spacing between these levels \cite{Gopal1966specific}. 
For $U>-1$ (where the F phase is found at low temperatures), we have found the typical $\lambda$-like singularity behavior of specific heat capacity, which increases its maximum value with the increasing $L$ as well as the rapid change in sublattice magnetizations from $|m|=0$ to $|m|=1$ with the decreasing temperature. 
To distinguish the order of these phase transitions, we have used the method by Challa et al. \cite{ChallaPRB1986,BinderRPP1987,farky2010,farky2014}, which is based on the energy distribution functions $P(E)$ for temperatures near phase transition.
For $U$ smaller than $U \approx -0.65$ (temperatures smaller than $k_BT \approx 0.2$), the energy distribution function exhibits a two-peaked structure near the phase transition (as shown in Fig.~\ref{fig:propferathf} for $U = -0.7$ and $k_{B}T \approx 0.19$), indicating a first-order phase transition from the NO phase to the F phase with decreasing temperature. 
This double-peak distribution indicates that states with two different energies of the system are equally probable at the transition temperature and one of them corresponds to the metastable solution in the vicinity of the transition line (however, the double-peak structure resembles a wide peak for model parameters shown).
On the other hand, for higher $U$, only one peak in $P(E)$ is observed, which denotes a second-order phase transition (absence of metastability in the system) as shown for $U = -0.6$ and $k_{B}T \approx 0.28$.
We also presents the results for $U=5.0$, where we get the second-order transition at $k_{B}T \approx 0.604$, which is in very good accordance with the exact result for the Ising model ($k_{B}T \approx 0.607$) \cite{HoutappelPhys1950a,HoutappelPhys1950b}.
However, the fluctuations are more visible due to higher temperatures and more Monte Carlo steps would provide smoother curves.

\begin{figure*}[htb]
\includegraphics[width=\sizeone]{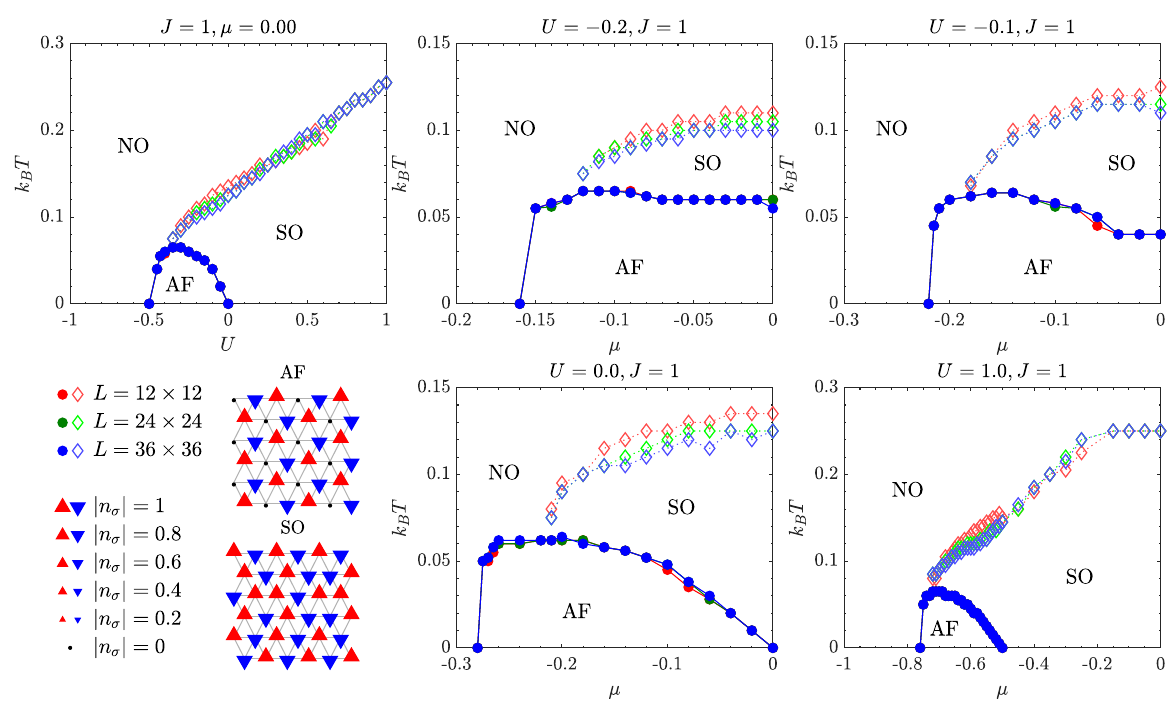}
\caption{\label{fig:PDantiferro}%
Phase diagrams for the model with antiferromagnetic ($J=1$) interactions, shown for three different lattice sizes ($L = 12^{2},24^{2}, 36^{2}$). Circular markers represent second-order phase transitions and diamonds illustrates the crossover from short-range order (SO) to non-ordered phase (NO). 
In the bottom-left corner, Monte Carlo averaged spin configurations for the identified antiferromagnetic phase (AF) at $U=1$, $\mu=-0.6$, and $k_BT=0.02$, and for the short-ordered (SO) phase at $U=1$, $\mu=-0.3$, and $k_BT=0.02$, are shown for a lattice of size $L = 6 \times 6$.}
\end{figure*}

Now we are ready to analyze the behavior of the system with changing $\mu$ (for $n\neq 1$).
Exemplary $k_{B}T$--$\mu$ phase diagrams for different values of $U$ are shown in Fig.~\ref{fig:PDferro}.
For all values of $U$ one sees the universal behavior of F--NO transition  temperature, namely, the increase of $|\mu|$ leads to a decrease of the temperature.
In the case of $U=-0.7$, the transition remains of first-order across the entire range of the parameter $\mu$. 
Another behavior is observed for $U$ values larger than $U \approx -0.65$, where a second-order transition occurs near $\mu=0$ and, with increasing $|\mu|$, changes its order into a first-order transition (cf. diagrams for $U=-0.6$ and $U=-0.5$). 
With further increasing of onsite repulsion $U$, the region of the F phase occurrence extends and the first-order line becomes almost vertical (cf. $U=0.0$ case).
It is remarkable that the temperature at which the F-NO transition changes its order (the tricritical point) is almost independent of $U$ and $\mu$, remaining around $k_{B}T \approx 0.18 - 0.20$, which is in an agreement with previous studies \cite{KapciaPhysA2015,KapciaJMMM2024}.

In Fig.~\ref{fig:propferawayhf}, we present dependence of several quantities with increasing temperature away half-filling to give an insight and justify the phase diagrams discussed above. 
The first column ($U=-0.7$, $\mu=-0.15$) presents the dependencies in the NO phase with the broad peak in the specific heat capacity accompanied by zero sublattice magnetizations and continuous decrease in number of particles with decreasing temperature. 
The second and the third columns ($U-0.7$, $\mu=-0.10$ and $U=-0.6$, $\mu=-0.18$) clearly indicate that, with increasing temperature, the first-order F-NO transition occurs, with a sharp (discontinuous) change of $m$ and $n$ at the transition temperature, and double peak-structure in the energy distribution $P(E)$ near the transition (which is more clearly visible than previously discussed example in Fig.~\ref{fig:propferathf}).
Finally, it is clearly seen that for $U=-0.6$ and $\mu=-0.12$ (the fifth column) the F-NO transition is second-order.
Note that the particle concentration $n$ approaches zero in the NO phase at the ground state, whereas in the F phase it goes to $1$ as it is expected for the exact ground-state results \cite{KapciaPhysA2015,KapciaJMMM2024}.

It is worth to mention that the findings for $J=-1$ are in a very good agreement, at least qualitatively, with the results obtained by the mean-field approach presented in \cite{KapciaPhysA2015,KapciaJMMM2024}. 
Obviously, the mean-field overestimates the transition temperatures.
Moreover, the results for the ferromagnetic coupling are also very similar (qualitatively and quantitatively) to those obtained for the square lattice \cite{MurawskiAPPA2012,MurawskiAPPA2014,MurawskiAPPA2015,KapciaPhysA2015}.
The predicted transition temperatures for the triangular lattice ($z=6$) are slightly higher than those for the square lattice ($z=4$), which is due to the higher number of the nearest-neighbors in the former case.

\subsection{Antiferromagnetic interactions}
\label{sec:antiferro}

\begin{figure*}[htb]
\includegraphics[width=\sizeone]{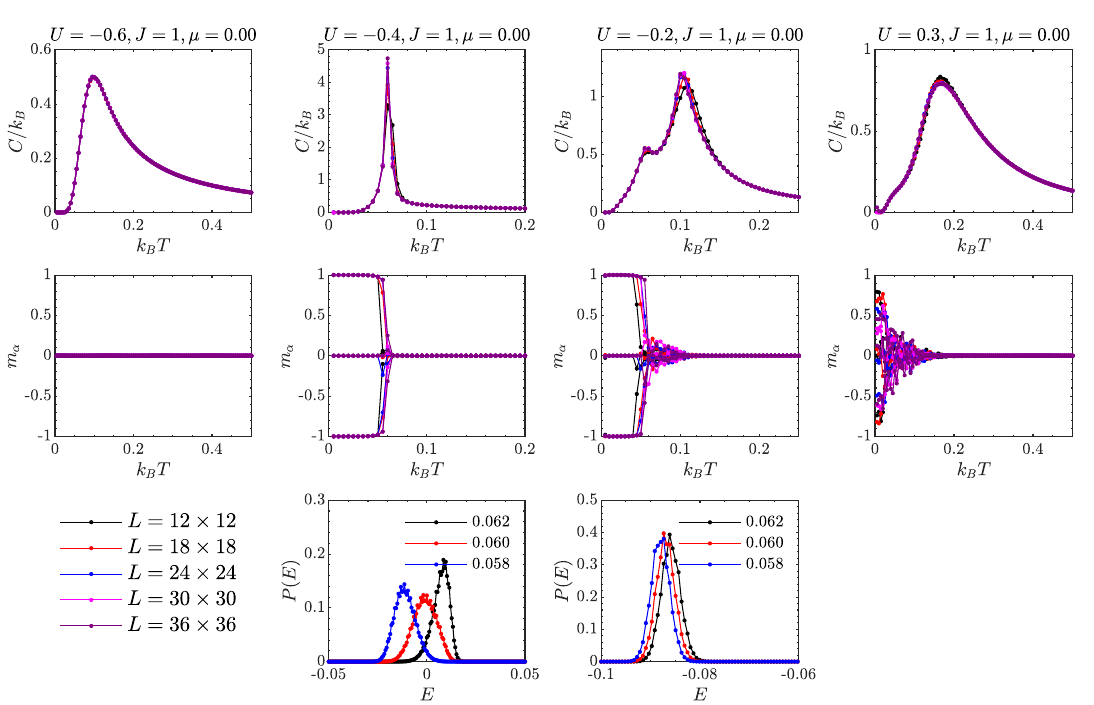}
\caption{\label{fig:propantiferroathf}%
Specific heat capacities (first line), sublattice magnetizations (second line), and the energy distribution functions near phase transition on a $24 \times 24$ lattice (third line) for antiferromagnetic interaction ($J=1$) and representative values of $U$ ($U =-0.6,-0.4, -0.3, 0.3$, from the left) at $\mu=0$.}
\end{figure*}

The case of antiferromagnetic interaction $J = 1$ (see Fig.~\ref{fig:PDantiferro}) is more complex, with a phase transition occurring only from the long-range antiferromagnetic (AF) phase, which, for $\mu=0$, occurs from $U = -0.5$ to $U = 0$ (see $k_{B}T$-$U$ phase diagram for $\mu=0$ in Fig.~\ref{fig:PDantiferro}). 
This AF phase is not an antiferromagnet in an usual meaning.
In the AF phase, we found an alternate arrangement of magnetic moments in two sublattices (e.g., $A$ and $B$ sublattices, which form the hexagonal lattice; $m_A=-m_B$), whereas in the third sublattice (e.g., sublattice $C$) an average magnetization is zero ($m_C = 0$, the direction of magnetic moment in a single realization is random), see snapshots in Fig.~\ref{fig:PDantiferro} (the labels of sublattices can be interchanged cyclically, cf. \cite{KapciaNano2021,KapciaJMMM2024}).
In particular, for the half-filling case ($\mu=0$), this AF phase consists of the $C$ sublattice being a combination of empty and double-occupied sites (with compensated magnetic moments) occurring with equal probability.

On the phase diagram, we also identified the SO state, which exhibits only short-range ordering. 
Compared to mean-field results \cite{KapciaPhysA2015,KapciaJMMM2024}, the spin configurations in this phase have similar combinations of $(\uparrow,\uparrow,\downarrow)$ and $(\uparrow,\downarrow,\downarrow)$ blocks. 
In contrary to the AF-NO transition line, the absence of long-range ordering does not lead to occurrence of phase transitions and the crossover from the SO state to the NO phase is found.
It is consistent with a fact that the SO state is not a new phase and it is just a state inside the NO phase. 
However, we explicitly distinguish the region of the SO state occurrence on the phase diagrams for $J>0$.

Similarly to the previous case, the position of the phase transitions and the crossover lines has been determined based on specific heat capacity and sublattice magnetizations, which are shown for representative values of $U$ in Fig.~\ref{fig:propantiferroathf} (for $\mu=0$). 
The first column shows the typical behavior of the NO phase (for $U=-0.6$), characterized by zero sublattice magnetizations and a specific heat capacity curve that is not scaled with $L$. 
For $U \approx -0.4$, we observed the typical $\lambda$ singularity in the specific heat capacity curve, and the energy distribution functions indicate a second-order phase transition. 
Increasing $U$ gives rise to another peak in specific heat capacity, which is represented on the plot for $U=-0.2$. 
One can see that only the low-temperature peak scales (i.e., goes to infinity) with increasing lattice size $L$. 
At this temperature ($k_{B}T\approx0.06$), we also observe a rapid change in sublattice magnetizations, which reveals the presence of a phase transition of the second order.
On the other hand, the high-temperature peak (at $k_{B}T \approx 0.10$ for $U=-0.2$) shows a broad maximum with a slight shift as $L$ increases. 
Between these two peaks we observe fluctuations of sublattice magnetization, which are caused by the presence of the short-range ordering (and antiferromagnetic interactions causing frustration in the system).
Similar behavior at the crossover from the non-ordered (NO) to short-range order (SO) phase with decreasing temperature is found also for $U>0$ (presented for $ U = 0.3$).

Let us now proceed to the discussion of effects of nonzero chemical potential (a case of finite particle doping) and its influence on the behavior of the system, which is illustrated for representative values of $U$ in the $k_BT-\mu$ phase diagrams shown in Fig.~\ref{fig:PDantiferro}.
It should be noted that the antiferromagnetic AF phase occurring for $\mu \neq 0$ preferred empty sites in the $C$ sublattice for $\mu<0$ ($n<1$) and double-occupied sites for $\mu>0$ ($n>1$). 
Similar to $\mu=0$, the phase transitions occur only if the AF phase is present in the ground state. 
An increase in the chemical potential $|\mu|$ leads to an almost constant transition temperature for $U\approx-0.2$, or, alternatively, to initial increase of the transition temperature near $\mu \approx 0$ for $U=-0.1$.
It drops to zero at the edge of the AF phase for $U=0$.
The maximal AF-NO transition temperature moves to $\mu \neq 0 $ with increasing $U$.
Conversely, increasing $|\mu|$ lowers monotonously the temperature of the crossover between the NO phase and SO state for all values of $U$.
Note also that, for $\mu\neq 0$ and $U=1$ the AF phase occurs in finite temperatures even it does not exist for $\mu = 0$ (such cases occur for $U>0$).
It is also worth to mention that the maximal temperature of the AF phase occurrence is almost not dependent on $\mu$ and $U$ and it equals $k_BT \approx 0.06$, approximately.

\begin{figure*}[bth]
\includegraphics[width=\sizetwo]{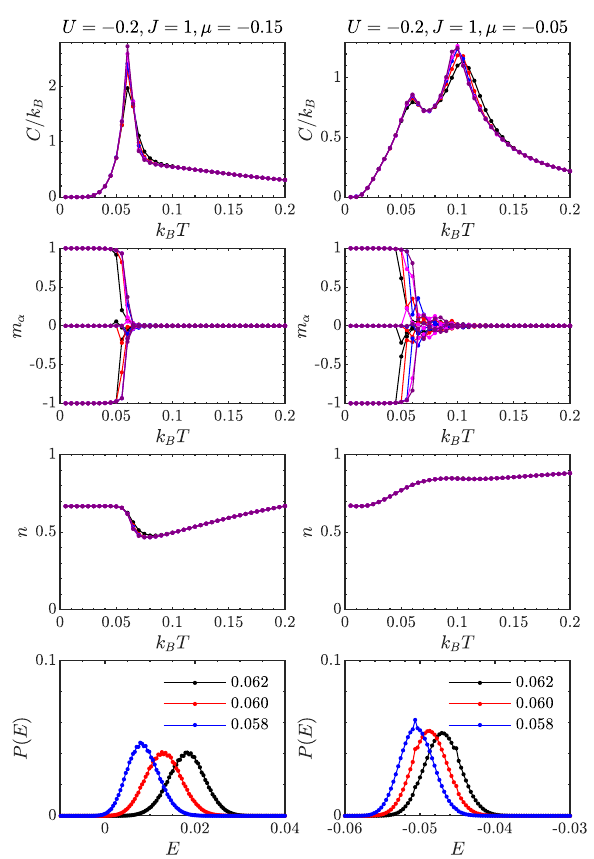}
\includegraphics[width=\sizetwo]{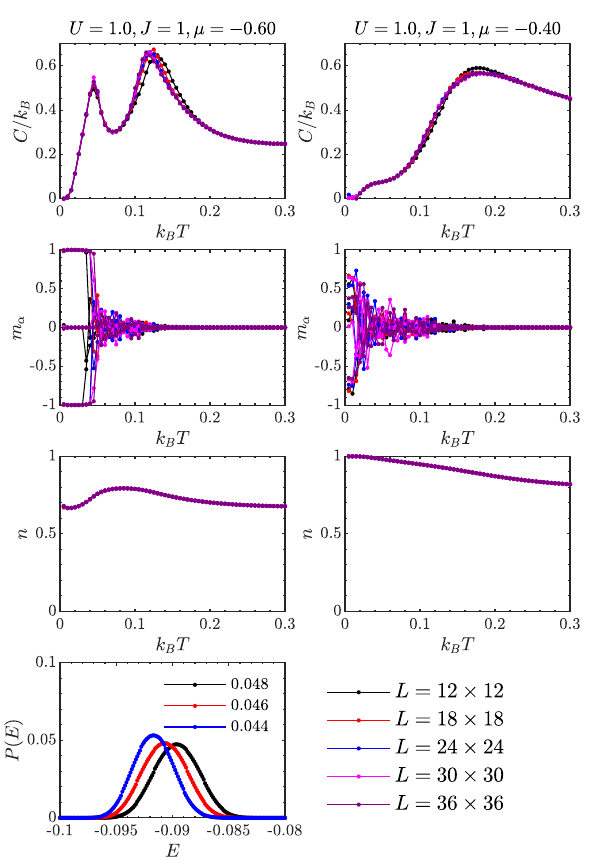}
\caption{\label{fig:propantiferroawayhf}%
Specific heat capacities (first line), sublattice magnetizations (second line), number of particles (third row), and the energy distribution functions near phase transition on a $24 \times 24$ lattice (fourth line) for antiferromagnetic interaction ($J=1$) and representative values of $U$ and $\mu$ (from the left: $U = -0.2$ and 
$\mu = -0.15, -0.05$; 
$U = 1.0$ and 
$\mu = -0.60,-0.40$).}
\end{figure*} 

Analogously to the previous case, in Fig.~\ref{fig:propantiferroawayhf}, we illustrate several physical quantities as a function of the temperature for representative values of $U$ ($U=-0.2$ and $U=1.0$) and $\mu$, which are used to construct the phase diagrams discussed above. 
The first column (for $U=-0.2$ and $\mu = -0.15$) shows the direct phase transition from the AF to the NO phase, characterized by a single peak in the specific heat capacity curve, a rapid change in sublattice magnetizations, and a number of particles approaching $n=2/3$ (and concentrations in the sublattices $n_A=n_B = 1$, $n_C = 0$) at the ground state. 
The second and third columns (for $U=-0.2$ and $\mu=-0.05$ or $U=1.0$ and $\mu=-0.60$) show specific heat capacity curves with two peaks: the low-temperature peak corresponds to the phase transition from the AF phase to the SO state (inside the NO phase), while the high-temperature peak corresponds to the crossover between the SO state and the NO phases. 
It can be observed that the energy distribution function near the AF-SO transition temperatures exhibits a single peak, indicating that the phase transitions are of second order.  
The last column (for $U=1.0$ and $ \mu=-0.40$) illustrates the case with the short-ordered ground state, where $n \rightarrow 1$ if $k_BT \rightarrow 0$ and sublattice magnetizations fluctuate strongly up to the crossover to the NO phase.

One should underline that we preformed detailed analysis of the energy distribution $P(E)$ for various model parameters near the transition from the AF phase (also for $L = 60 \times 60 $ system and smaller steps of $\Delta (k_{B}T)=0.002,0.001$ and $0.0002$ for a few sets of the parameters) and we do not find any indicators for the first-order nature of the transition from the AF phase. 
We also decreased the intervals of $E$ to calculate $P(E)$ distribution, but two-peak structure of $P(E)$ did not appeared. 
Thus, we are quite confident that the AF--NO and AF--SO transitions are second-order (continuous) ones.

\section{Conclusion}
\label{sec:conclusions}

In this work we investigated the localized fermion system on a triangular lattice described by the extended Hubbard model with onsite Coulomb $U$ and nearest-neighbor Ising-like magnetic $J$ interactions by means of the Monte carlo simulations in the grand canonical ensemble.
The investigated model (\ref{eq:hamiltonian}) is a non-trivial extension of the well-known Ising model.
By examining various quantities, such as specific heat capacity, sublattice magnetizations, number of particles and energy distribution functions, we construct comprehensive phase diagrams for both signs of $J$ interaction across varying temperatures and values of $U$ and chemical potential ($\mu$).

In the ferromagnetic case ($J < 0$), our results confirm the presence of the phase transition from the ferromagnetic (F) phase to the non-ordered (NO) phase. Interestingly, the transition order depends on the values of $U$ and $\mu$. 
For temperatures lower than $k_{B} T \approx 0.18-0.20$, it is first-order, whereas for higher temperatures, the transition becomes second-order. 
The highest temperature for the F phase occurrence $k_{B}T \approx 0.605$ is found for $U\rightarrow +\infty$ and $\mu=0$ (in the limit, the model reduces to the Ising model).
These findings are consistent with the mean-field results as well as with results obtained using Monte Carlo simulations for the square lattice, albeit with differences in transition temperature magnitudes (due to different number of nearest neighbors).

The results for antiferromagnetic case ($J>0$) are very distinct from that obtained for $J<0$ case.
This is because, in this case, the model exhibits geometrical frustration and thus the findings are strongly affected by the frustration.
The region of the occurrence of the ordered AF phase is strongly reduced with only second-order transition from the AF phase (which is not an usual antiferromagnet).
For $\mu = 0$,  the AF phase exist only for $U$ in the range of $-0.5<U<0$.
For $U>0$ the AF phase can occur only for $\mu\neq 0$. 
The maximal temperature for the AF phase occurrence is $k_{B}T \approx 0.06$ and it is almost independent on $U$ or $\mu$.
Note also that two sublattices, from which single occupied sites form the antiferromagnetic pattern in the AF phase defined here (with $n\approx 2/3$ or $4/3$), create the honeycomb lattice. 
For the Ising model on the honeycomb lattice ($z=3$), the critical temperature for the ordered phase is found as $kT_c/(2|J|) = 0.253$  \cite{HoutappelPhys1950b}.

One should underline that our results coincide with the rigorous results at the ground state \cite{KapciaJMMM2024} and some limiting cases ($U\rightarrow+\infty$, $\mu=0$) \cite{HoutappelPhys1950a,HoutappelPhys1950b}. 
In addition, our distinct approach reveals intriguing finite-temperature behaviors and phase transitions, providing better understanding of the effects of competing interactions in strongly frustrated systems.
Although model (\ref{eq:hamiltonian}) is relatively simple and can be treated as a toy model, it represents a non-trivial generalization of the Ising model, offering valuable insights into phase behavior in strongly correlated systems. These findings not only deepen our understanding of such systems but also provide a foundation for future research into related magnetic and electronic lattice models. 
In a case of longer-range interactions in the system, new phases exhibiting also various incommensurate ordering can occur \cite{KaburagiJPSJ1978,BakPRL1982,BakRPP1982,KerimovJMP2020}.

Finally, one should note that the considered system described by hamiltonian (\ref{eq:hamiltonian}) is not as far from the reality as one can expect. 
In particular, the arrangement of magnetic moments in the triangular lattice has been observed in various materials, e.g., in  
Li$_{2}$MnTeO$_{6}$ \cite{ZverevaPRB2020},
KCeO$_{2}$ \cite{BordelonPRB2021},
KCeS$_{2}$ \cite{KulbakovJPCM2021},
Ba$_{2}$MnTeO$_{6}$ \cite{KhatuaSciRep2021,LiPRB2022},
Ce$_{3}$Cu and Pr$_{3}$Cu \cite{OgunbunmiJAC2022},
Co$_{1/3}$TaS$_2$ \cite{ParkNatComm2023}, as well as in CsCeSe$_{2}$~\cite{XiePRB2024}.
Moreover, also ultra-cold atomic gases give the opportunity of realization of systems with different geometries \cite{PartridgeScience2006,ZwierleinScience2006,SchunckScience2007}. 
Because these systems have  controllable parameters, they are perfect to study fundamental phenomena and allow to simulate different models in various regimes of parameters \cite{BlochRMP2008,GiorginiRMP2008,GeorgescuRMP2014,DuttaRPP2015,SchaferNatRevPhys2020tools,JinFrontPhys2022}.
The geometry of the lattice can be experimentally changed by different spatial arrangement of laser beams.
In particular, it is possible to create the triangular lattice \cite{BeckerNJP2010,YamamotoNJP2020,WangPRL2023}.
As neglecting of the single-particle hopping might be considered as an
artificial one, it can be recognized as the first step for description of the magnetism in the mentioned systems.
These facts create the future opportunity of the experimental validation of the predictions presented in the current work.

\begin{acknowledgments}
We thank Dr. Pavol Farka\v{s}ovsk\'y and Dr. Jan Bara\'nski for very fruitful discussions and careful reading of the manuscript.
L.~R. thanks to Slovak Research and Development
Agency under the contract no. APVV-20-0293 and the Slovak Grant 
Agency Vega under the contract no. 2/0037/22.
K.~J.~K. thanks the Polish National Agency for Academic Exchange for funding in the frame of the National Component of the Mieczysław Bekker program (2020 edition) (BPN/BKK/2022/1/00011).
L.~R. acknowledges hospitality and research stays in Department of Theory of Condensed Matter (Adam Mickiewicz University).
\end{acknowledgments}

\bibliography{biblio}

\end{document}